# Universal pacemaker of genome evolution

Sagi Snir[1], Yuri I. Wolf[2], Eugene V. Koonin[2,*]

[1] - Department of Evolutionary and Environmental Biology and The Institute of Evolution, University of Haifa Mount Carmel, Haifa 31905, Israel.
[2] - National Center for Biotechnology Information, NLM, National Institutes of Health, Bethesda, Maryland 20894, USA

[*]Corresponding author.

**Molecular clock (MC) is a central concept of molecular evolution according to which each gene evolves at a characteristic, near constant rate[1-3]. Numerous evolutionary studies have demonstrated the validity of MC but also have shown that MC is substantially overdispersed, i.e. lineage-specific deviations of the evolutionary rate of the given gene from the clock greatly exceed the expectation from the sampling error[4-6]. A fundamental observation of comparative genomics that appears to complement the MC is that the distribution of evolution rates across orthologous genes in pairs of related genomes remains virtually unchanged throughout the evolution of life, from bacteria to mammals[7-9]. The conservation of this distribution implies that the relative evolution rates of all genes remain nearly constant, or in other words, that evolutionary rates of different genes are strongly correlated within each evolving genome. We hypothesized that this correlation is not a simple consequence of MC but could be better explained by a model we dubbed Universal PaceMaker (UPM) of genome evolution. The UPM model posits that the rate of evolution changes synchronously across genome-wide sets of genes in all evolving lineages. We sought to differentiate between the MC and UPM models by fitting thousands of phylogenetic trees for bacterial and archaeal genes to supertrees that reflect the dominant trend of vertical descent in the evolution of archaea and bacteria and that were constrained according to the two models. The goodness of fit for the UPM model was better than the fit for the MC model, with overwhelming statistical significance. These results reveal a universal pacemaker of genome evolution that could have been in operation throughout the history of life.**

In 1962 Zuckerkandl and Pauling discovered that the number of differences between homologous proteins is roughly proportional to the divergence time separating the corresponding species[10,11]. This phenomenon became known as 'molecular clock' (MC) and has been supported by multiple independent observations[3,12]. The MC is the basis of molecular dating whereby the age of an evolutionary event, usually the split between lineages (such as for example humans and chimpanzee), is estimated from the sequence divergence using calibration with dates known from fossil record[13-15]. From the phylogenetic point of view, when genes evolve along a rooted tree under the MC, branch lengths are proportional to the time between speciation (or duplication) events and the distances from each internal tree node to all descendant leaves are the same (ultrametric tree) up to the precision of the estimation (the latter being determined by sampling error which is inevitable in comparison of finite-length sequences).

Over the 50 years that elapsed since the seminal finding of Zuckerkandl and Pauling, several important amendments to the MC concept have been made. First, MC has been shown to be overdispersed, i.e. the differences between the root to tip distances in many or most subtrees of a given tree usually greatly exceed the expectation from sampling error[4-6]. This observation has led to the development of the relaxed MC model which is a compromise between an unconstrained tree with arbitrary branch lengths and an MC tree[16,17]. Under the relaxed MC, the evolutionary rate is allowed to change from branch to branch but this change is presumed to be gradual so that related lineages evolve at similar rates. The relaxed MC model underlies most of the modern methods of molecular dating. Second, different genes within the same genome have been found to evolve at different rates within the range of three to four orders of magnitude[8,9,18].

The strict MC implies that all orthologous genes present in a group of organisms and sharing the same evolutionary history evolve in a fully coherent manner even if at different rates.

Indeed, if the divergence between gene sequences is solely determined by the divergence time and gene-specific evolution rate, phylogenetic trees reconstructed from different genes will have the same topology and nearly identical branch lengths up to a scaling factor which is equal to the relative evolution rate. Under the MC model, the differences between the corresponding branch lengths in different gene trees are due solely to the sampling error which arises from stochastic factors and is expected to be uncorrelated between trees. The relaxed MC model allows greater, non-random deviations in the lengths of corresponding branches but to our knowledge, the possibility that these evolution rate changes are correlated between genes has not been considered explicitly.

Genome-wide analysis of distances between orthologous genes in pairs of organisms from a broad range of taxa belonging to all three domains of life (bacteria, archaea and eukaryotes) revealed striking similarity between the distributions of these distances. All these distributions are approximately lognormal and are nearly identical in shape, up to a scaling factor[7,9]. Although many different explanations are possible of this remarkable conservation of evolutionary rate distribution across the entire spectrum of life, the simplest underlying model is that all genes evolve at approximately constant rates relative to each other. In other words, evolution of genes seems to be strongly correlated genome-wide, with changes in the rate of evolution occurring in unison.

The constancy of gene-specific relative evolution rate does not imply MC. Although MC is sufficient to produce the same effect, it is not necessary. The deviations of the absolute evolution rates from the clock could be arbitrarily high but, if they apply to all genes in the genome to the same degree, the relative evolutionary rates would remain approximately the same throughout the entire course of evolution and in all lineages.

The idea that the apparent constancy of relative evolution rates across genome-wide gene sets is caused by synchronous changes in the evolution of (nearly) all genes in the genome can be embodied in the model of Universal PaceMaker (UPM) of genome evolution. While the strict MC ticks at a constant rate for all genes at all times and in all lineages, the UPM can arbitrarily change its rate relative to the astronomical clock at any time, but the corresponding acceleration or deceleration of the evolution rate applies to all genes in the genome. Here we compare the MC and UPM models of evolution by analysis of phylogenetic trees for a genome-wide set of prokaryotic gene families and show that the UPM model is a better fit for the evolution of prokaryotes.

Our data set consisted of the "forest" of phylogenetic trees reconstructed for 6901 orthologous gene families representing 41 archaeal and 59 bacterial genomes[19] (see Supplementary Material). Although horizontal gene transfer is widespread in the evolution of prokaryotes[20,21], the tree-like statistical trend is detectable in the genome-wide data set and moreover dominates the evolution of (nearly) ubiquitous gene families[19,22]. We encapsulate this trend in a rooted supertree (ST) that reflects the prevalent vertical descent in the evolution of archaea and bacteria (see Supplementary Material). Each individual original gene tree (GT) is compared to the ST and reduced to the maximum agreement subtree (MAST), i.e. the largest set of leaves whose phylogeny fits the ST topology. Removal of discordant nodes and edges leads to collapse of several edges of the original GT into a single edge (Figure 1); then, the length of the newly created GT edge is the sum of the original contributing GT edges. Likewise, when GT is mapped to ST, several adjacent ST edges could correspond to a single edge in the reduced GT, forming a composite edge.

Under both the MC and the UPM models, we assume that the lengths of the ST edges determine the expected lengths of the corresponding GT edges. For the MC model, edge lengths correspond to time intervals between speciation events, the ST is strictly ultrametric, and gene-specific evolutionary rates are measured in substitutions per site per time unit. Under the UPM model, edge lengths represent arbitrarily defined "ticks" of the universal pacemaker (internal time), and gene-specific evolutionary rates are measured in substitutions per site per pacemaker unit of internal time. More formally,

$$l_{i,k} = t_j r_k \varepsilon_{i,k}$$

where $l_{i,k}$ is the length of the $i$-th edge of the $k$-th GT, $t_j$ is length of the $j$-th (possibly composite) ST edge corresponding to the $i$-th edge of the $k$-th GT, $r_k$ is the gene-specific evolution rate, and $\varepsilon_{i,k}$ is the multiplicative error factor for the given edge. We further assume that the error is random, independent for branches both within and between GTs, and comes from a lognormal distribution with the mean of 1 and an arbitrary variance, translating to a model with an additive normally distributed deviation in the logarithmic scale. Because the distributions of evolutionary rates tend to follow symmetric bell-shaped curves in log scale [9,23], the assumption of a multiplicative, log-normally distributed deviation seems natural.

First, we seek to find the set of ST edge lengths ***t*** and gene rates ***r*** that provides the best fit to the entire set of GTs. Under the assumption of a normally distributed deviation, the likelihood function for the set of GTs given ***t*** and ***r*** is

$$\ln L(\boldsymbol{t}, \boldsymbol{r}) \approx -\frac{n}{2}(\ln E^2 - \ln n + \ln 2\pi + 1)$$

where $n$ is the total number of edges in the set of GTs and $E^2$ is the sum of squares of deviations between the expected and observed edge lengths in the logarithmic scale:

$$E^2 = \sum_k E_k^2 = \sum_k \sum_i (\ln l_{i,k} - \ln t_j r_k)^2$$

where the summation for $i$ is done over the edges of a given GT and the summation for $k$ is done over all GTs (see Supplementary Material). Thus, finding the maximum likelihood solution for $\{\boldsymbol{t}, \boldsymbol{r}\}$ is equivalent to finding the minimum of $E^2$. For the MC model, the ST edge lengths $\boldsymbol{t}$ are constrained by the ultrametricity requirement, whereas for the UPM model, ST edge lengths are unconstrained.

For the analyzed set of 100 genomes, there is a choice of several possible ST topologies, produced using different methods (see Supplementary Material). We mapped all original GTs onto each of these STs and obtained reduced GTs that corresponded to the respective MASTs. The GTs that yielded MASTs with fewer than 10 leaves were discarded. The ST topology derived from the concatenated alignments of ribosomal proteins provided the maximum total number of leaves in the resulting set of reduced GTs and accordingly was chosen for further analysis. Altogether, we obtained 2294 reduced GTs with MAST size greater or equal to 10 species including 44,889 leaves and 82,896 edges. This set of trees was fit to an ultrametricity-constrained ST (MC model) and an unconstrained ST (UPM model) (Table 1, see Supplementary Material).

We then compared the MC and UPM models in terms of the goodness of fit to the data. Obviously, the residual sum of squares is lower for the UPM model because it involves independent optimization of all 198 ST edge lengths, whereas under the MC model the edge lengths are subject to 99 ultrametricity constraints. To account for the difference in the numbers of degrees of freedom, we employed the Akaike Information Criterion (AIC) to compare the MC and UPM models. Under the assumption of normally distributed deviations:

$$\Delta AIC = AIC_{MC} - AIC_{UPM} = n \ln \frac{E_{MC}^2}{E_{PM}^2} + 2\Delta d$$

where $E^2_{MC}$ and $E^2_{UPM}$ are the residual sums of squares for the MC and UPM models, respectively, $n$ is the total number of GT edges and $\Delta d$ is the difference in the number of parameters optimized in the process of fitting (in our case $\Delta d = -99$). Because lower AIC values correspond to better quality of fit, negative $\Delta AIC$ would indicate preference for the MC model whereas a positive $\Delta AIC$ would indicate support for the UPM model. The relative likelihood weight of the suboptimal model can be estimated as $1/\exp(|\Delta AIC|/2)$.

The results presented in Table 1 reveal overwhelming support of the UPM model over the MC model. Thus, evolutionary rates tend to change synchronously for the majority (if not all) of the genes in evolving genomes although the rate of the UPM relative to the astronomical time differs for different lineages. The results of this analysis show that the apparent genome-wide constancy of the relative rates of gene evolution across vast spans of life's history (Figure 2A) is not a trivial consequence of MC but in part results from a distinct, fundamental evolutionary phenomenon, the UPM (Figure 2B). It seems a natural possibility that UPM is instigated by shifts in population dynamics of evolving lineages with changes affecting all genes in the same manner and to a similar degree.

The difference between the UPM and MC models is highly significant but small in magnitude. Root mean square deviation (r.m.s.d.) of GT edges from the expectations derived from UMP ST is large (a factor of 2.45) and only slightly less that the r.m.s.d for the MC ST (a factor of 2.48). Thus, similar to MC, the UPM appears to be substantially overdispersed. To assess the robustness of the finding that UPM fits the GTs better than MC, we isolated the contributions of individual trees to the $E^2_{MC}$ and $E^2_{UPM}$ ($E^2_{MC,k}$ and $E^2_{UPM,k}$ respectively), took 1000 bootstrap samples of the set of GTs and computed $\Delta AIC$ values for each sample. All 1000 $\Delta AIC$ values obtained for the resampled sets were positive (in the range of 1511 to 2147), providing 100% support to the superiority of the UPM model and ensuring that this result is consistent for the majority of the GTs and is not determined by a small number of strongly biased trees (see Supplementary Material for details). The distribution of the $E^2_{MC,k}$ / $E^2_{UPM,k}$ ratio (Figure 3A) shows a strong bias toward values greater than unity (73% of the GTs), supporting the robustness of this result.

The $E^2_{MC,k}$ / $E^2_{UPM,k}$ ratio characterizes the degree to which the *k*-th GT favors the UPM model. Linear model analysis shows that this value is significantly and independently influenced by the overall goodness of fit to the ST (p-value <<0.001), the fraction of the original GT leaves remaining in the MAST with ST (p-value <<0.001) and the number of the original GT leaves (p-value <<0.001). Thus, the GTs that retain higher number of leaves in MAST, fit the ST better and are wider distributed among prokaryotes, typically show a stronger preference for the UPM model over the MC model. These three factors together explain ~9% of the variance in $\ln(E^2_{MC,k} / E^2_{UPM,k})$. Neither the relative evolution rate nor the functional class of the gene significantly impact the degree of preference of UPM over MC (see Supplementary Material for details). Interpreting these findings in terms closer to biology, widely-distributed genes that are subject to relatively little horizontal transfer or sporadic changes of evolution rate that reduce the fit to ST appear to make the greatest contribution to the UPM. These observations imply that the UPM is indeed a fundamental feature of genome evolution, at least in prokaryotes.

The distribution of estimated relative evolution rates (Figure 3B) spans values within a range of slightly greater than an order of magnitude (0.26 to 4.58). This is considerably less than the range of rates measured over short evolutionary distances [9,23]. Accelerations and decelerations of the UPM might average out over long intervals of evolution, reducing the observed differences between genes.

The discovery of the UPM opens up several areas of further inquiry. In particular, it remains to be determined whether or not distinct pacemakers govern the evolution of different classes of genes. The biological connotations of the UPM are of major interest. Mapping UPM shifts to specific stages of the evolution of life, changes in the life style and population structure of organisms as well as to the geological record could become an important direction of future research.

## Methods

### Supertrees and Maximum Agreement Subtrees

Three distinct supertrees (STs) were tested for the purpose of representing the vertical inheritance trend in the analyzed set of GTs. The first supertree ($ST_1$) was from [19] (originally computed using the CLANN program [24]; the second supertree ($ST_2$) was computed using the quartet supertree method [25] for all species quartets in the complete set of GTs the third supertree ($ST_3$) was derived from a tree of concatenated sequences of (nearly) universal ribosomal proteins [26]. Maximum Agreement Subtrees (MAST) between the supertree (ST) and any given gene tree (GT) were computed using the *agree* program of the PAUP* package [27]. The set of MASTs with the analyzed GTs was computed for each of these STs, yielding a total of 43,068 MAST leaves for $ST_1$, 43,411 MAST leaves for $ST_2$ and 44,889 MAST leaves for $ST_3$ (MAST $\geq$10 for each ST). Accordingly, $ST_3$ was used for all further analyses as the topology that best represented the entire set of GTs.

To perform the LS optimization of the ST edge lengths and the GT relative evolution rates, we used the function *fmin_slsqp()* that is part of the *scipy.optimize* package of Python which minimizes a function using sequential least squares programming. The function also adopts a set of constraints that are necessary for the calculation. In both the MC and the UPM models, both the ST edges and the GT rates were constrained to positive values. For the UPM model, the distances from a node to any leaf in a subtree under that node were set equal for all subtrees. It can be shown by induction that this constraint implies an ultrametric tree. Thus, we have a constraint for every internal node; in a rooted binary tree with $m$ leaves, there are $m$ - 1 such nodes.

**Acknowledgements**

We thank Pere Puigbo and Natalya Yutin for generous help with handling the phylogenetic trees. SS was supported by a grant from the U.S.-Israel Binational Science Foundation. YIW and EVK were supported by the Department of Health and Human Services intramural program (NIH, National Library of Medicine).

## Figure Legends

### Figure 1. Gene trees and the supertree

A. A gene tree (GT). After the comparison with the supertree (ST), the GT is reduced to the maximum agreement subtree (MAST, highlighted in yellow). The reduced GT edge highlighted in red corresponds to two edges in the original GT.

B. Supertree (ST). Mapping of the reduced GT onto the ST is highlighted; two sections of ST that consist of multiple edges mapping to a single edge of the reduced GT are highlighted in blue and green, respectively.

### Figure 2. Distributions of relative rates of gene evolution

A. The distribution of the $E^2_{MC,k}$ / $E^2_{UPM,k}$ ratios for 2294 gene families

B. The $r_k$ value distribution for 2294 gene families was obtained by fitting gene trees to the UPM supertree

The distribution curves were smoothed using the Gaussian-kernel method

### Figure 3. The Universal Molecular Clock and Universal Pacemaker models of genome evolution

A. Under the Molecular Clock model, gene-specific evolution rates (colored lines) remain constant; at any point in time (shown as dots), the relative rates of gene evolution are also constant.

B. Under the Universal Pacemaker model, gene-specific evolution rates can change arbitrarily but by the same amount across the entire genome; at any point in time, the relative rates of gene evolution remain constant.

**Table 1. Comparison of the Molecular Clock and Universal Pacemaker models of genome evolution**

| Model of genome evolution | MC | UPM |
|---|---|---|
| **Number of trees** | 2,294 | 2,294 |
| **Number of leaves** | 44,889 | 44,889 |
| **Number of edges** | 82,896 | 82,896 |
| $E^2$ | 68,260.8 | 66,626.5 |
| **r.m.s.d., ln units** | 0.9074 | 0.8965 |
| **r.m.s.d., factor** | 2.4780 | 2.4510 |
| $\Delta AIC$ | 1810.8 | 0 |
| **Relative likelihood weight** | $10^{-393}$ | 0 |

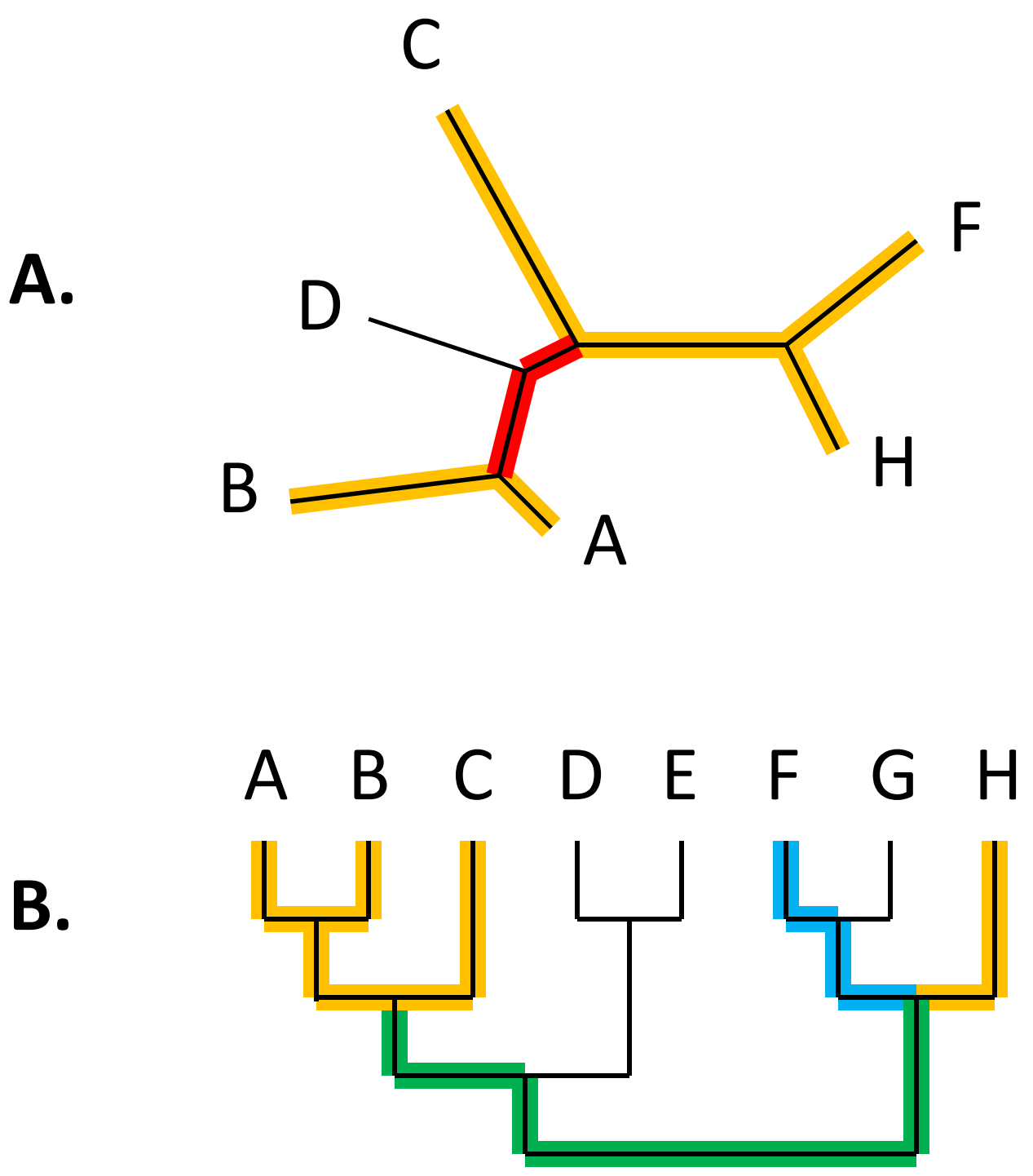

Figure 1

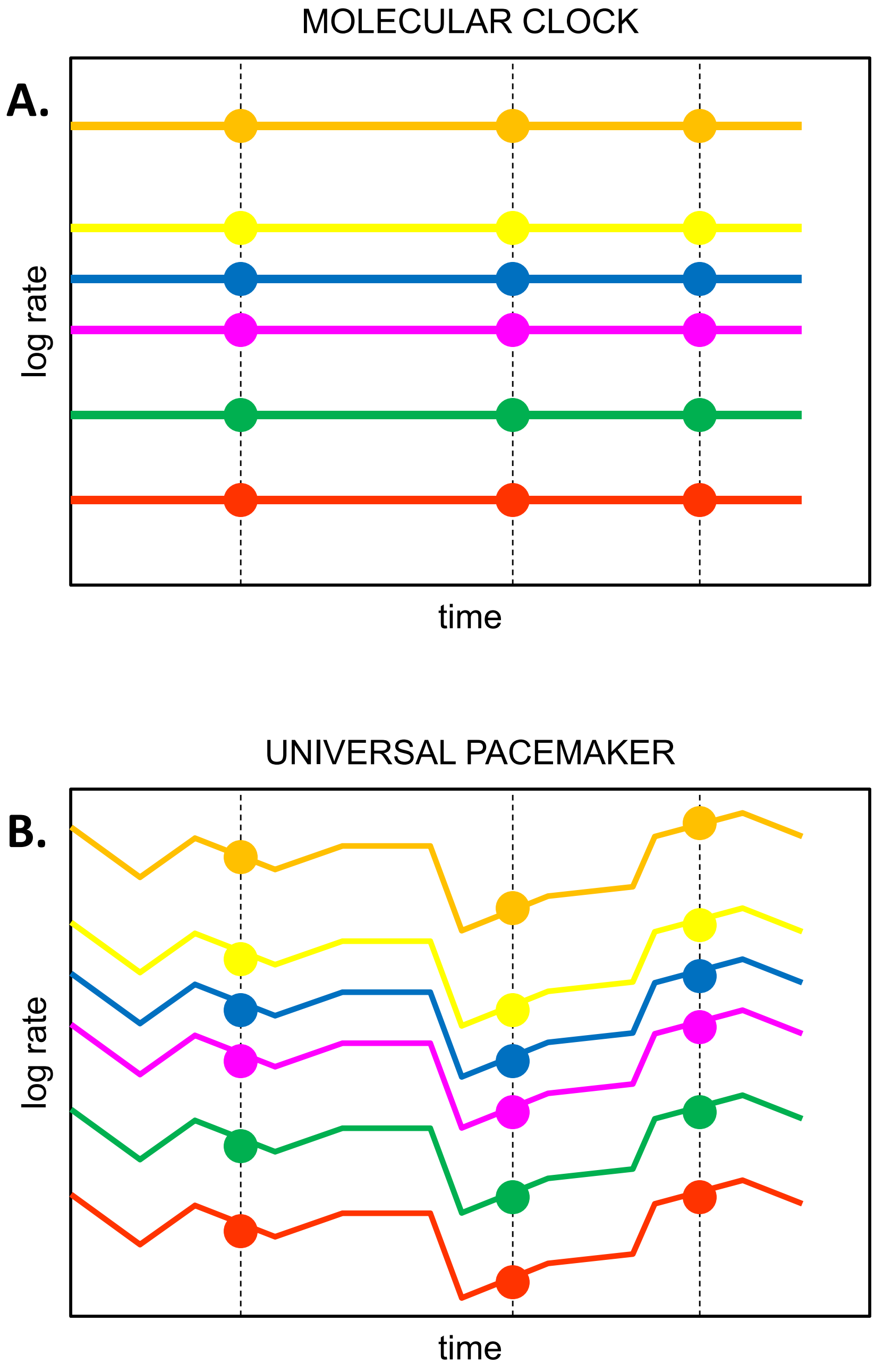

Figure 2

**A.**

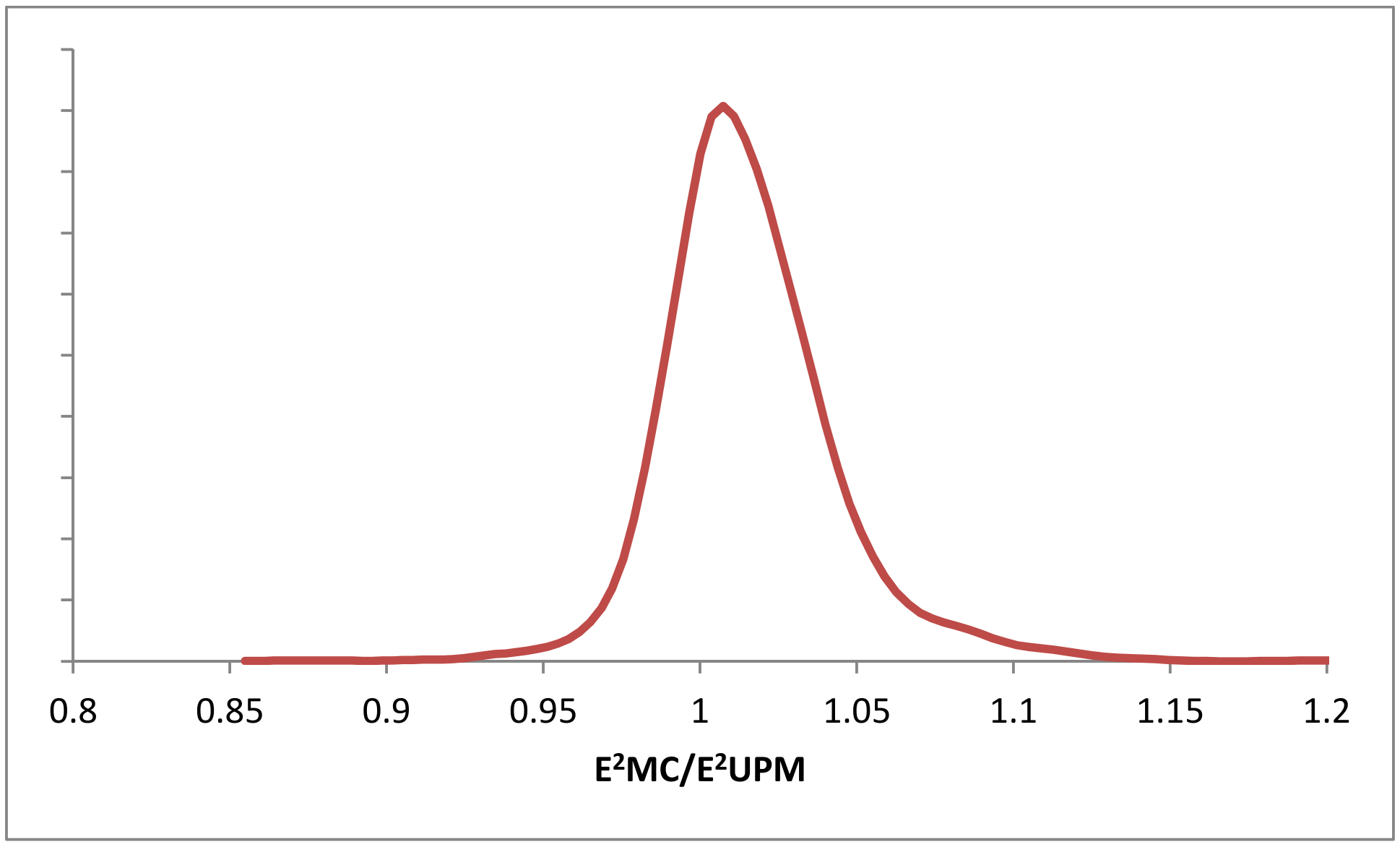

**B.**

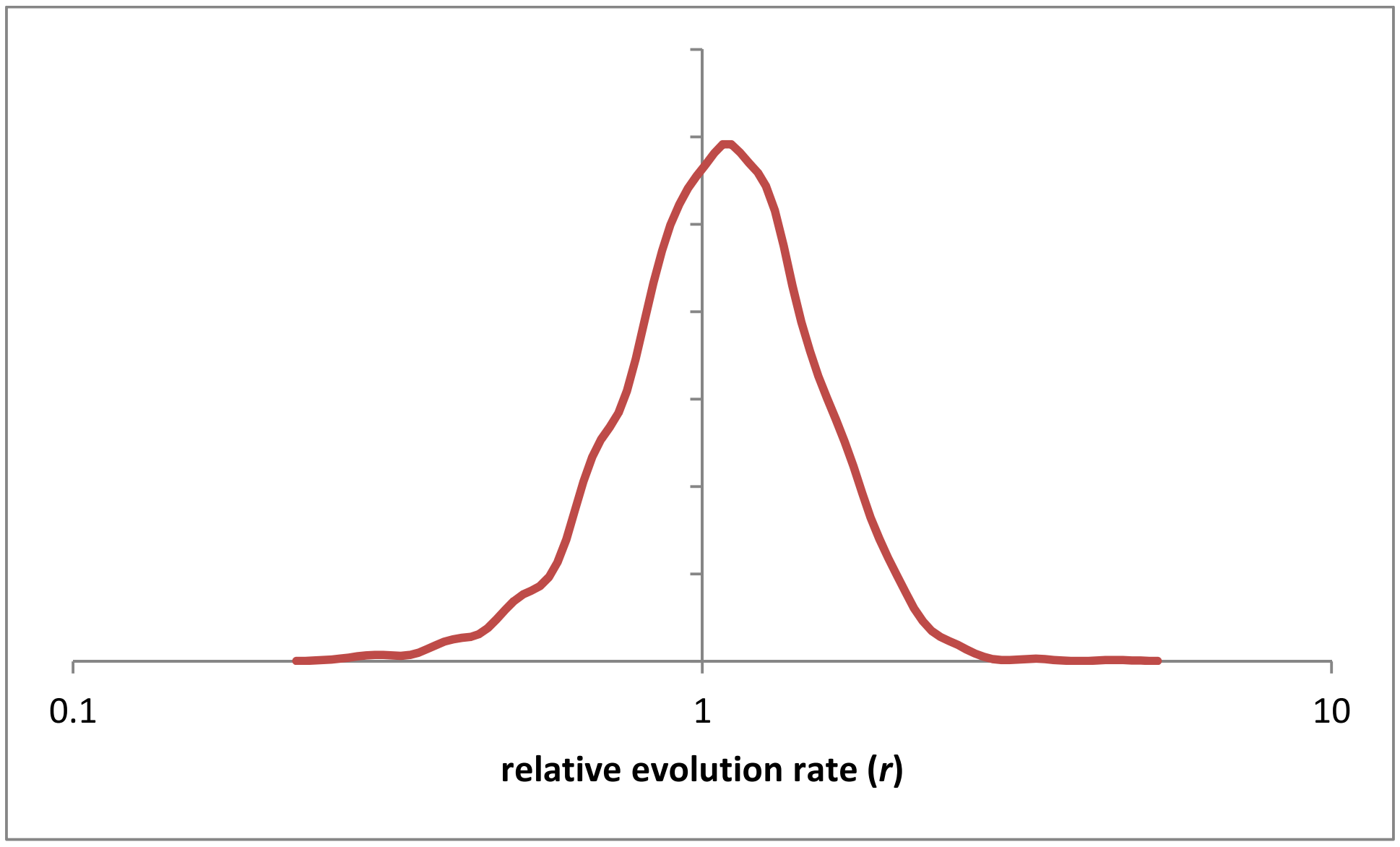

0.1
1
10
relative evolution rate (*r*)

Figure 3

# Supplementary Material

## Universal pacemaker of genome evolution

Sagi Snir[1], Yuri I. Wolf[2], Eugene V. Koonin[2,*]

**Supertree ($ST_3$) topology.**

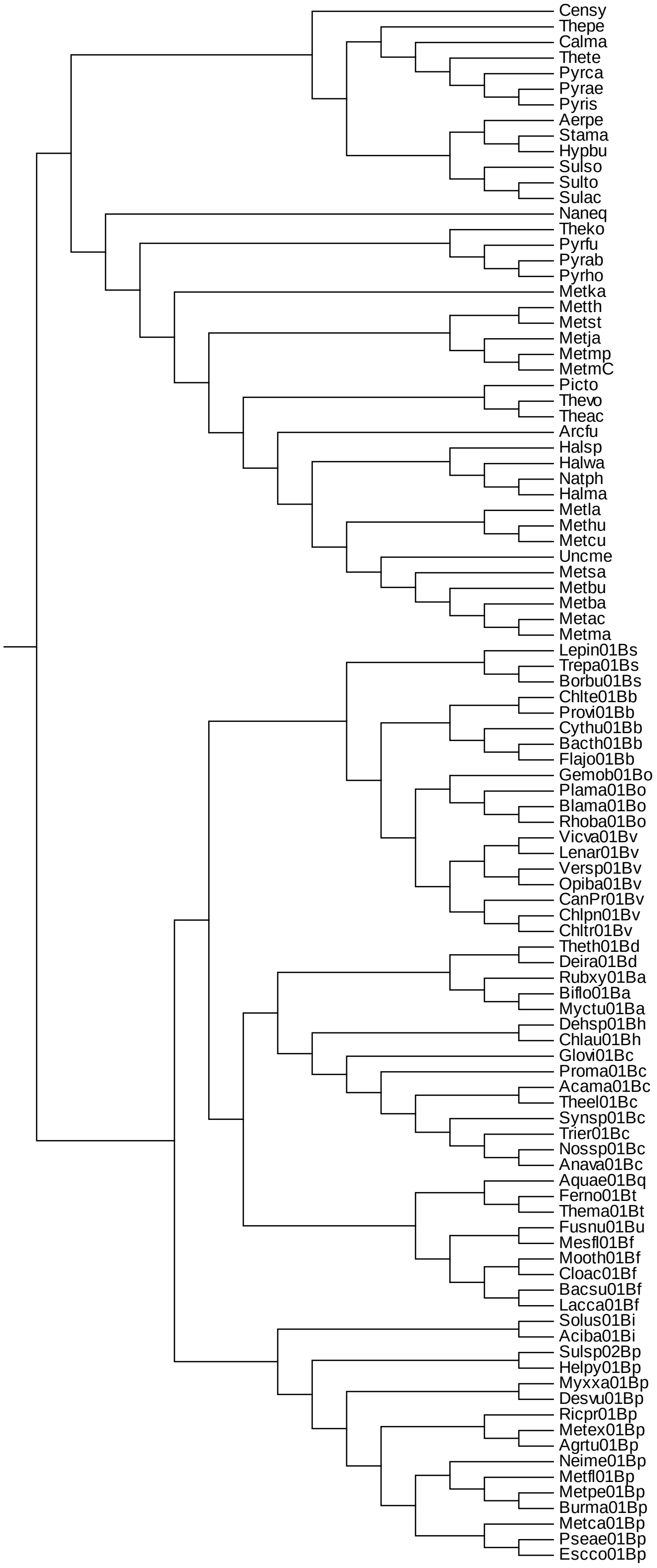

## Supertree ($ST_3$) topology, Newick format:

(((Censy,((Thepe,(Calma,(Thete,(Pyrca,(Pyrae,Pyris))))),((Aerpe,(Stama,Hypbu)),(Sulso,(Sulto,Sulac))))),(Naneq,((Theko,(Pyrfu,(Pyrab,Pyrho))),(Metka,(((Metth,Metst),(Metja,(Metmp,MetmC))),((Picto,(Thevo,Theac)),(Arcfu,((Halsp,(Halwa,(Natph,Halma))),((Metla,(Methu,Metcu)),(Uncme,(Metsa,(Metbu,(Metba,(Metac,Metma)))))))))))))),((((Lepin01Bs,(Trepa01Bs,Borbu01Bs)),(((Chlte01Bb,Provi01Bb),(Cythu01Bb,(Bacth01Bb,Flajo01Bb))),((Gemob01Bo,(Plama01Bo,(Blama01Bo,Rhoba01Bo))),(((Vicva01Bv,Lenar01Bv),(Versp01Bv,Opiba01Bv)),(CanPr01Bv,(Chlpn01Bv,Chltr01Bv)))))),((((Theth01Bd,Deira01Bd),(Rubxy01Ba,(Biflo01Ba,Myctu01Ba))),((Dehsp01Bh,Chlau01Bh),(Glovi01Bc,(Proma01Bc,((Acama01Bc,Theel01Bc),(Synsp01Bc,(Trier01Bc,(Nossp01Bc,Anava01Bc)))))))),((Aquae01Bq,(Ferno01Bt,Thema01Bt)),((Fusnu01Bu,Mesfl01Bf),((Mooth01Bf,Cloac01Bf),(Bacsu01Bf,Lacca01Bf)))))),((Solus01Bi,Aciba01Bi),((Sulsp02Bp,Helpy01Bp),((Myxxa01Bp,Desvu01Bp),((Ricpr01Bp,(Metex01Bp,Agrtu01Bp)),((Neime01Bp,(Metfl01Bp,(Metpe01Bp,Burma01Bp))),(Metca01Bp,(Pseae01Bp,Escco01Bp)))))))));

**List of species in the dataset.**

| | | | | |
|---|---|---|---|---|
| Censy | *Cenarchaeum symbiosum* | Archaea | Crenarchaeota | Cenarchaeales |
| Aerpe | *Aeropyrum pernix* | Archaea | Crenarchaeota | Desulfurococcales |
| Hypbu | *Hyperthermus butylicus* | Archaea | Crenarchaeota | Desulfurococcales |
| Stama | *Staphylothermus marinus F1* | Archaea | Crenarchaeota | Desulfurococcales |
| Sulac | *Sulfolobus acidocaldarius DSM 639* | Archaea | Crenarchaeota | Sulfolobales |
| Sulso | *Sulfolobus solfataricus* | Archaea | Crenarchaeota | Sulfolobales |
| Sulto | *Sulfolobus tokodaii* | Archaea | Crenarchaeota | Sulfolobales |
| Calma | *Caldivirga maquilingensis IC-167* | Archaea | Crenarchaeota | Thermoproteales |
| Pyrae | *Pyrobaculum aerophilum* | Archaea | Crenarchaeota | Thermoproteales |
| Pyrca | *Pyrobaculum calidifontis JCM 11548* | Archaea | Crenarchaeota | Thermoproteales |
| Pyris | *Pyrobaculum islandicum DSM 4184* | Archaea | Crenarchaeota | Thermoproteales |
| Thepe | *Thermofilum pendens Hrk 5* | Archaea | Crenarchaeota | Thermoproteales |
| Thete | *Thermoproteus tenax* | Archaea | Crenarchaeota | Thermoproteales |
| Uncme | *Uncultured methanogenic archaeon* | Archaea | Euryarchaeota | ? |
| Arcfu | *Archaeoglobus fulgidus* | Archaea | Euryarchaeota | Archaeoglobales |
| Halma | *Haloarcula marismortui ATCC 43049* | Archaea | Euryarchaeota | Halobacteriales |
| Halsp | *Halobacterium sp* | Archaea | Euryarchaeota | Halobacteriales |
| Halwa | *Haloquadratum walsbyi* | Archaea | Euryarchaeota | Halobacteriales |
| Natph | *Natronomonas pharaonis* | Archaea | Euryarchaeota | Halobacteriales |
| Metth | *Methanobacterium thermoautotrophicum* | Archaea | Euryarchaeota | Methanobacteriales |
| Metst | *Methanosphaera stadtmanae* | Archaea | Euryarchaeota | Methanobacteriales |
| Metja | *Methanococcus jannaschii* | Archaea | Euryarchaeota | Methanococcales |
| MetmC | *Methanococcus maripaludis C5* | Archaea | Euryarchaeota | Methanococcales |
| Metmp | *Methanococcus maripaludis S2* | Archaea | Euryarchaeota | Methanococcales |
| Metla | *Methanocorpusculum labreanum Z* | Archaea | Euryarchaeota | Methanomicrobiales |
| Metcu | *Methanoculleus marisnigri JR1* | Archaea | Euryarchaeota | Methanomicrobiales |
| Methu | *Methanospirillum hungatei JF-1* | Archaea | Euryarchaeota | Methanomicrobiales |
| Metka | *Methanopyrus kandleri* | Archaea | Euryarchaeota | Methanopyrales |
| Metbu | *Methanococcoides burtonii DSM 6242* | Archaea | Euryarchaeota | Methanosarcinales |
| Metsa | *Methanosaeta thermophila PT* | Archaea | Euryarchaeota | Methanosarcinales |
| Metac | *Methanosarcina acetivorans* | Archaea | Euryarchaeota | Methanosarcinales |
| Metba | *Methanosarcina barkeri fusaro* | Archaea | Euryarchaeota | Methanosarcinales |
| Metma | *Methanosarcina mazei* | Archaea | Euryarchaeota | Methanosarcinales |
| Pyrab | *Pyrococcus abyssi* | Archaea | Euryarchaeota | Thermococcales |
| Pyrfu | *Pyrococcus furiosus* | Archaea | Euryarchaeota | Thermococcales |
| Pyrho | *Pyrococcus horikoshii* | Archaea | Euryarchaeota | Thermococcales |
| Theko | *Thermococcus kodakaraensis KOD1* | Archaea | Euryarchaeota | Thermococcales |
| Picto | *Picrophilus torridus DSM 9790* | Archaea | Euryarchaeota | Thermoplasmales |
| Theac | *Thermoplasma acidophilum* | Archaea | Euryarchaeota | Thermoplasmales |
| Thevo | *Thermoplasma volcanium* | Archaea | Euryarchaeota | Thermoplasmales |
| Naneq | *Nanoarchaeum equitans* | Archaea | Nanoarchaeota | ? |
| Aciba01Bi | *Acidobacteria bacterium Ellin345* | Bacteria | Acidobacteria | Acidobacteriales |
| Solus01Bi | *Solibacter usitatus Ellin6076* | Bacteria | Acidobacteria | Solibacterales |
| Myctu01Ba | *Mycobacterium tuberculosis H37Rv* | Bacteria | Actinobacteria | Actinomycetales |
| Biflo01Ba | *Bifidobacterium longum NCC2705* | Bacteria | Actinobacteria | Bifidobacteriales |
| Rubxy01Ba | *Rubrobacter xylanophilus DSM 9941* | Bacteria | Actinobacteria | Rubrobacterales |
| Aquae01Bq | *Aquifex aeolicus VF5* | Bacteria | Aquificae | Aquificales |
| Bacth01Bb | *Bacteroides thetaiotaomicron VPI-5482* | Bacteria | Bacteroidetes | Bacteroidales |
| Flajo01Bb | *Flavobacterium johnsoniae UW101* | Bacteria | Bacteroidetes | Flavobacteriales |
| Cythu01Bb | *Cytophaga hutchinsonii ATCC 33406* | Bacteria | Bacteroidetes | Sphingobacteriales |
| CanPr01Bv | *Candidatus Protochlamydia amoebophila UWE25* | Bacteria | Chlamydiae | Chlamydiales |
| Chltr01Bv | *Chlamydia trachomatis D/UW-3/CX* | Bacteria | Chlamydiae | Chlamydiales |
| Chlpn01Bv | *Chlamydophila pneumoniae AR39* | Bacteria | Chlamydiae | Chlamydiales |
| Chlte01Bb | *Chlorobium tepidum TLS* | Bacteria | Chlorobi | Chlorobiales |
| Provi01Bb | *Prosthecochloris vibrioformis DSM 265* | Bacteria | Chlorobi | Chlorobiales |
| Chlau01Bh | *Chloroflexus aurantiacus J-10-fl* | Bacteria | Chloroflexi | Chloroflexales |
| Dehsp01Bh | *Dehalococcoides sp BAV1* | Bacteria | Chloroflexi | Dehalococcoidetes |
| Synsp01Bc | *Synechocystis sp PCC 6803* | Bacteria | Cyanobacteria | Chroococcales |
| Theel01Bc | *Thermosynechococcus elongatus BP-1* | Bacteria | Cyanobacteria | Chroococcales |
| Glovi01Bc | *Gloeobacter violaceus PCC 7421* | Bacteria | Cyanobacteria | Gloeobacterales |
| Anava01Bc | *Anabaena variabilis ATCC 29413* | Bacteria | Cyanobacteria | Nostocales |
| Nossp01Bc | *Nostoc sp PCC 7120* | Bacteria | Cyanobacteria | Nostocales |
| Trier01Bc | *Trichodesmium erythraeum IMS101* | Bacteria | Cyanobacteria | Oscillatoriales |
| Proma01Bc | *Prochlorococcus marinus subsp marinus str CCMP1375* | Bacteria | Cyanobacteria | Prochlorales |
| Acama01Bc | *Acaryochloris marina MBIC11017* | Bacteria | Cyanobacteria | unclassified |
| Deira01Bd | *Deinococcus radiodurans R1* | Bacteria | Deinococci | Deinococcales |
| Theth01Bd | *Thermus thermophilus HB27* | Bacteria | Deinococci | Thermales |
| Bacsu01Bf | *Bacillus subtilis subsp subtilis str 168* | Bacteria | Firmicutes | Bacillales |
| Cloac01Bf | *Clostridium acetobutylicum ATCC 824* | Bacteria | Firmicutes | Clostridiales |
| Mesfl01Bf | *Mesoplasma florum L1* | Bacteria | Firmicutes | Entomoplasmatales |
| Lacca01Bf | *Lactobacillus casei ATCC 334* | Bacteria | Firmicutes | Lactobacillales |
| Mooth01Bf | *Moorella thermoacetica ATCC 39073* | Bacteria | Firmicutes | Thermoanaerobacteriales |
| Fusnu01Bu | *Fusobacterium nucleatum subsp nucleatum ATCC 25586* | Bacteria | Fusobacteria | Fusobacteriales |
| Lenar01Bv | *Lentisphaera araneosa HTCC2155* | Bacteria | Lentisphaerae | Lentisphaerales |
| Vicva01Bv | *Victivallis vadensis ATCC BAA-548* | Bacteria | Lentisphaerae | Victivallales |
| Blama01Bo | *Blastopirellula marina DSM 3645* | Bacteria | Planctomycetes | Planctomycetales |
| Gemob01Bo | *Gemmata obscuriglobus* | Bacteria | Planctomycetes | Planctomycetales |
| Plama01Bo | *Planctomyces maris DSM 8797* | Bacteria | Planctomycetes | Planctomycetales |
| Rhoba01Bo | *Rhodopirellula baltica SH 1* | Bacteria | Planctomycetes | Planctomycetales |
| Agrtu01Bp | *Agrobacterium tumefaciens str C58* | Bacteria | Proteobacteria-Alpha | Rhizobiales |
| Metex01Bp | *Methylobacterium extorquens PA1* | Bacteria | Proteobacteria-Alpha | Rhizobiales |
| Ricpr01Bp | *Rickettsia prowazekii str Madrid E* | Bacteria | Proteobacteria-Alpha | Rickettsiales |
| Burma01Bp | *Burkholderia mallei ATCC 23344* | Bacteria | Proteobacteria-Beta | Burkholderiales |
| Metpe01Bp | *Methylibium petroleiphilum PM1* | Bacteria | Proteobacteria-Beta | Burkholderiales |
| Metfl01Bp | *Methylobacillus flagellatus KT* | Bacteria | Proteobacteria-Beta | Methylophilales |
| Neime01Bp | *Neisseria meningitidis MC58* | Bacteria | Proteobacteria-Beta | Neisseriales |
| Desvu01Bp | *Desulfovibrio vulgaris subsp vulgaris str Hildenborough* | Bacteria | Proteobacteria-Delta | Desulfovibrionales |
| Myxxa01Bp | *Myxococcus xanthus DK 1622* | Bacteria | Proteobacteria-Delta | Myxococcales |
| Helpy01Bp | *Helicobacter pylori 26695* | Bacteria | Proteobacteria-Epsilon | Campylobacterales |
| Sulsp02Bp | *Sulfurovum sp NBC37-1* | Bacteria | Proteobacteria-Epsilon | unclassified |
| Escco01Bp | *Escherichia coli K12* | Bacteria | Proteobacteria-Gamma | Enterobacteriales |
| Metca01Bp | *Methylococcus capsulatus str Bath* | Bacteria | Proteobacteria-Gamma | Methylococcales |
| Pseae01Bp | *Pseudomonas aeruginosa PAO1* | Bacteria | Proteobacteria-Gamma | Pseudomonadales |
| Borbu01Bs | *Borrelia burgdorferi B31* | Bacteria | Spirochaetes | Spirochaetales |
| Lepin01Bs | *Leptospira interrogans serovar Copenhageni str Fiocruz L1-130* | Bacteria | Spirochaetes | Spirochaetales |
| Trepa01Bs | *Treponema pallidum subsp pallidum str Nichols* | Bacteria | Spirochaetes | Spirochaetales |
| Ferno01Bt | *Fervidobacterium nodosum Rt17-B1* | Bacteria | Thermotogae | Thermotogales |
| Thema01Bt | *Thermotoga maritima MSB8* | Bacteria | Thermotogae | Thermotogales |
| Opiba01Bv | *Opitutaceae bacterium TAV2* | Bacteria | Verrucomicrobia | Opitutales |
| Versp01Bv | *Methylokorus infernorum V4* | Bacteria | Verrucomicrobia | Verrucomicrobiales |

**Maximum Likelihood estimate for the supertree edge lengths and gene evolution rates**

Consider a rooted supertree (ST) with a fixed topology. The ST encompasses a set of edges $\boldsymbol{e}$ defined by the ST topology and a set of unknown edge lengths $\boldsymbol{t}$. Consider a set of unrooted GTs reduced to MAST with the given ST. Each GT encompasses a set of edges with known edge lengths and an unknown gene-specific evolution rate ($\boldsymbol{b}_k$, $\boldsymbol{l}_k$ and $r_k$ for the $k$-th GT, respectively). Each edge of each GT uniquely maps to an ST path $\boldsymbol{e}_j$, that is a subset of adjacent edges in the ST ($b_{k,i} \equiv \boldsymbol{e}_j$ where $\boldsymbol{e}_j \subseteq \boldsymbol{e}$ for the $i$-th edge of the $k$-th GT).

Let $t_j = \sum_{x \in \boldsymbol{e}_j} t_x$ be the length of the path $\boldsymbol{e}_j$. We assume that the length of the $i$-th edge of the $k$-th GT is related to the length of the corresponding ST path $\boldsymbol{e}_j$:

$$l_{i,k} = t_j r_k \varepsilon_{i,k}$$

where $\varepsilon_{i,k}$ is the multiplicative deviation factor for the given edge. We further assume that the deviation is random, independent for branches both within and between GTs, and comes from a lognormal distribution with the mean of 1 and an arbitrary variance, translating to a model with an additive normally distributed deviation in the logarithmic scale (i.e. $\ln \varepsilon_{i,k} \sim N(0,\sigma^2)$).

Given $\boldsymbol{t}$ and $\boldsymbol{r}$, the expectation for the logarithm of the length of the $i$-th edge of the $k$-th GT is:

$$\mu_{i,k} = \langle \ln l_{i,k} \rangle = \langle \ln t_j \rangle + \langle \ln r_k \rangle + \langle \ln \varepsilon_{i,k} \rangle = \ln t_j + \ln r_k$$

and the likelihood of observing the length $l_{i,k}$ is:

$$\Pr\{l_{i,k} | \boldsymbol{t}, \boldsymbol{r}\} = \frac{1}{\sigma\sqrt{2\pi}} \exp\left(-\frac{\left(\ln l_{i,k} - \mu_{i,k}\right)^2}{2\sigma}\right) = \frac{1}{\sigma\sqrt{2\pi}} \exp\left(-\frac{\left(\ln l_{i,k} - \ln t_j - \ln r_k\right)^2}{2\sigma}\right) = \frac{1}{\sigma\sqrt{2\pi}} \exp\left(-\frac{E_{i,k}^2}{2\sigma}\right)$$

where $E^2_{i,k} = (\ln l_{i,k} - \ln t_j - \ln r_k)^2$. For all observed edge lengths in all GTs ($\boldsymbol{l}$), the likelihood function is

$$L(\boldsymbol{l}|\boldsymbol{t}, \boldsymbol{r}) = \prod_k \prod_i \Pr\{l_{i,k} | \boldsymbol{t}, \boldsymbol{r}\}$$

In the logarithmic scale:

$$\ln L(\boldsymbol{l}|\boldsymbol{t}, \boldsymbol{r}) = \sum_k \sum_i \ln \frac{1}{\sigma\sqrt{2\pi}} \exp\left(-\frac{E_{i,k}^2}{2\sigma}\right) = -\frac{n}{2} \ln \sigma^2 - \frac{n}{2} \ln 2\pi - \frac{\sum_k \sum_i E_{i,k}^2}{2\sigma^2}$$

where $n$ is the total number of GT edges ($n = \sum_k \sum_i 1$). Designating the residual sum of squares $E^2 = \sum_k \sum_i E_{i,k}^2$ and substituting the estimate for $\sigma^2$

$$\widehat{\sigma^2} = \frac{E^2}{n-1} \approx \frac{E^2}{n}$$

for large *n*, we obtain:

$$\ln L(\boldsymbol{l}|\boldsymbol{t},\boldsymbol{r}) \approx -\frac{n}{2}\ln\frac{E^2}{n} - \frac{n}{2}\ln 2\pi - \frac{n}{2}$$

Because *n* is constant for a given data set, finding the maximum of $L(\boldsymbol{l} \mid \boldsymbol{t},\boldsymbol{r})$ is equivalent to finding the minimum of $E^2$.

**Optimization procedure**

Least Squares (LS) is called linear if the residuals are linear for all unknowns. Linear LS can be represented in a matrix format which has a closed form solution (given that the columns of the matrix are linearly independent). However, our formulation requires taking logs over sums of unknowns in the case where a GT edge corresponds to a path in ST ( $\ln t_j = \ln \sum_{x \in \boldsymbol{e}_j} t_x$). Then, the problem becomes non-linear with respect to LS and can be solved only using numerical algorithms where the solution is obtained by iteratively refining the parameter values. This approach requires supplying initial values for the parameters. The goodness of the initial value estimation is critical for the convergence time of the iterative method and the risk of being trapped in local maximum points. We employed the following strategy for determining the initial values: For each ST edge, we computed the mean value of the sum over all GT edges that uniquely correspond to the given edge. Therefore, if we assign one gene a specific rate value (e.g. the length of some edge), we obtain initial rate values for all genes. It can be easily shown that, if there are no errors in rates (i.e. $\sigma^2 = 0$), the above procedure yields the accurate (ML) values for all unknowns.

**Optimization of the fit between STs and GTs with different MAST size thresholds**

| | **MAST≥30** | **MAST≥20** | **MAST≥10** |
|---|---|---|---|
| No. of GTs | 246 | 967 | 2,294 |
| No. of MAST GT leaves | 9,134 | 26,441 | 44,889 |
| No. of MAST GT edges | 17,530 | 49,981 | 82,896 |
| $E^2_{MC}$ / $E^2_{UPM}$ | 1.045 | 1.031 | 1.025 |
| $\Delta AIC$ | 573.0 | 1,310.8 | 1,810.8 |
| $\Delta BIC$ | -196.4 | 437.7 | 887.6 |

BIC, Bayesian Information Criterion which is more conservative than the AIC with regard to model complexity because of a heavier penalty for extra degrees of freedom. The conservative character of BIC is the reason why it prefers the MC model for MAST ≥ 30 (the smallest data set).

## MC and UPM optimization of the supertree branch lengths.

MC-constrained (ultrametric) supertree

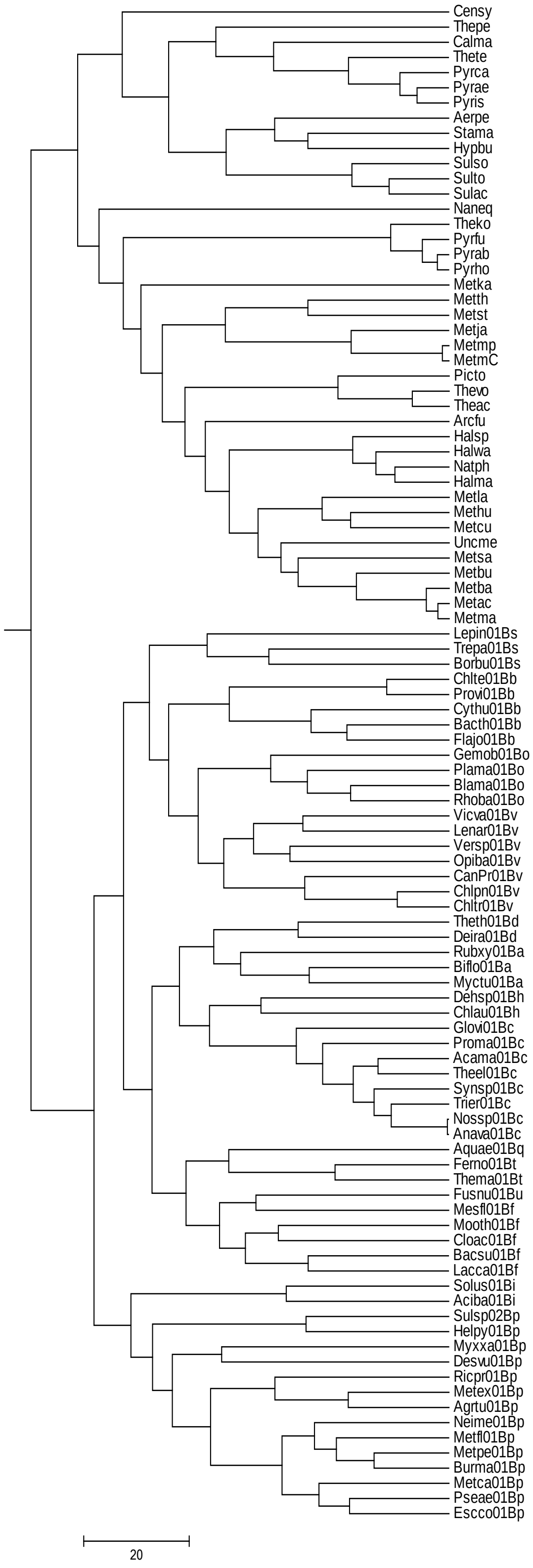

Unconstrained (UPM) supertree

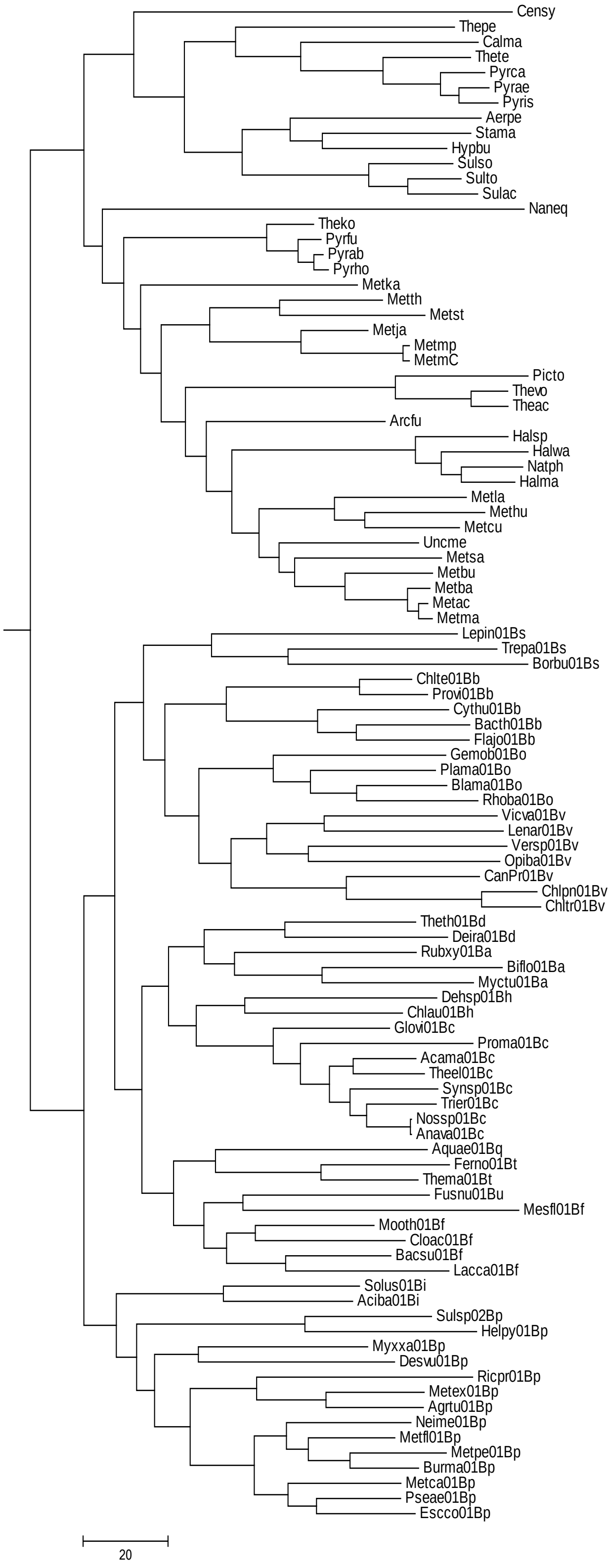

### MC-optimized supertree, Newick format.

(((Censy:61.8869,((Thepe:44.1285,(Calma:33.1595,(Thete:18.9522,(Pyrca:9.1778,(Pyrae:5.9390,Pyris:5.
9390):3.2388):9.7744):14.2072):10.9690):8.9634,((Aerpe:33.0161,(Stama:26.7236,Hypbu:26.7236):6.292
5):9.1865,(Sulso:18.3712,(Sulto:11.3010,Sulac:11.3010):7.0703):23.8314):10.8892):8.7951):8.4681,(Nan
eq:66.3254,((Theko:11.0283,(Pyrfu:5.0095,(Pyrab:2.1686,Pyrho:2.1686):2.8409):6.0188):50.7054,(Metka
:58.3823,(((Metth:26.7240,Metst:26.7240):15.6488,(Metja:18.5184,(Metmp:1.2279,MetmC:1.2279):17.29
05):23.8544):11.9734,((Picto:21.0763,(Thevo:6.9803,Theac:6.9803):14.0960):28.9880,(Arcfu:46.2026,((H
alsp:18.2850,(Halwa:13.8666,(Natph:10.2491,Halma:10.2491):3.6175):4.4184):23.3907,((Metla:24.0772,(
Methu:18.6710,Metcu:18.6710):5.4062):12.1627,(Uncme:31.9038,(Metsa:28.6607,(Metbu:17.6049,(Metb
a:4.3497,(Metac:2.1720,Metma:2.1720):2.1777):13.2552):11.0558):3.2431):4.3362):5.4357):4.5269):3.86
17):4.2820):4.0360):3.3514):4.5918):4.0296):8.8492,((((Lepin01Bs:45.8589,(Trepa01Bs:34.1513,Borbu0
1Bs:34.1513):11.7076):10.9256,(((Chlte01Bb:11.7560,Provi01Bb:11.7560):29.8532,(Cythu01Bb:26.1225,
(Bacth01Bb:19.3430,Flajo01Bb:19.3430):6.7795):15.4867):11.5212,((Gemob01Bo:33.7971,(Plama01Bo:
26.8322,(Blama01Bo:18.6302,Rhoba01Bo:18.6302):8.2020):6.9649):13.7445,(((Vicva01Bv:27.6823,Lena
r01Bv:27.6823):9.3475,(Versp01Bv:30.1312,Opiba01Bv:30.1312):6.8986):5.6725,(CanPr01Bv:27.3706,(
Chlpn01Bv:9.8091,Chltr01Bv:9.8091):17.5615):15.3316):4.8393):5.5889):3.6540):4.8775,((((Theth01Bd:2
8.5267,Deira01Bd:28.5267):16.0194,(Rubxy01Ba:39.4497,(Biflo01Ba:26.4593,Myctu01Ba:26.4593):12.9
904):5.0964):6.5052,((Dehsp01Bh:35.5921,Chlau01Bh:35.5921):9.7385,(Glovi01Bc:28.8885,(Proma01Bc
:23.8659,((Acama01Bc:13.3682,Theel01Bc:13.3682):4.7123,(Synsp01Bc:14.1482,(Trier01Bc:10.8841,(N
ossp01Bc:0.2118,Anava01Bc:0.2118):10.6723):3.2641):3.9323):5.7853):5.0227):16.4421):5.7207):5.194
8,((Aquae01Bq:41.6984,(Ferno01Bt:21.5559,Thema01Bt:21.5559):20.1425):8.0988,((Fusnu01Bu:36.535
3,Mesfl01Bf:36.5353):6.8947,((Mooth01Bf:32.2905,Cloac01Bf:32.2905):6.2299,(Bacsu01Bf:26.5997,Lacc
a01Bf:26.5997):11.9207):4.9097):6.3671):6.4489):5.4159):5.6010,((Solus01Bi:30.8122,Aciba01Bi:30.812
2):29.4297,((Sulsp02Bp:27.0372,Helpy01Bp:27.0372):29.0993,((Myxxa01Bp:43.0730,Desvu01Bp:43.073
0):9.3309,((Ricpr01Bp:32.9890,(Metex01Bp:19.0928,Agrtu01Bp:19.0928):13.8962):12.1896,((Neime01B
p:25.4812,(Metfl01Bp:21.3187,(Metpe01Bp:14.1583,Burma01Bp:14.1583):7.1604):4.1624):6.1640,(Metc
a01Bp:24.6509,(Pseae01Bp:18.8580,Escco01Bp:18.8580):5.7928):6.9943):13.5334):7.2253):3.7327):4.1
054):7.0210):11.9412);

## UPM-optimized supertree, Newick format

(((Censy:89.0259,((Thepe:51.4890,(Calma:41.7386,(Thete:20.5655,(Pyrca:10.4465,(Pyrae:7.0525,Pyris:
9.1721):4.3686):13.5020):19.3299):15.3357):12.0650,((Aerpe:44.9282,(Stama:34.9028,Hypbu:29.2652):7
.5422):11.3161,(Sulso:19.7941,(Sulto:12.5176,Sulac:16.4295):9.1575):29.7845):13.5787):11.9140):11.69
70,(Naneq:99.2062,((Theko:11.0680,(Pyrfu:5.4038,(Pyrab:2.1754,Pyrho:3.4121):3.7090):7.3922):33.640
2,(Metka:50.9979,(((Metth:24.1644,Metst:34.2067):16.4281,(Metja:15.7372,(Metmp:1.5000,MetmC:1.467
7):23.9853):21.4609):11.3870,((Picto:31.1798,(Thevo:8.6051,Theac:8.6739):17.7942):49.3736,(Arcfu:42.
0494,((Halsp:21.7838,(Halwa:20.4254,(Natph:14.4531,Halma:12.6005):4.7011):6.1063):43.1505,((Metla:
31.0090,(Methu:28.2125,Metcu:22.2977):7.1172):17.7435,(Uncme:32.8120,(Metsa:34.5754,(Metbu:20.61
89,(Metba:5.2481,(Metac:2.2004,Metma:3.2066):2.6050):14.7119):11.8231):3.6558):4.7029):6.4194):5.9
909):4.8941):5.7240):4.8166):3.9425):5.0274):4.3704):12.6121,((((Lepin01Bs:57.8117,(Trepa01Bs:49.14
54,Borbu01Bs:56.3925):17.9829):16.0285,(((Chlte01Bb:12.2697,Provi01Bb:16.0112):31.3288,(Cythu01B
b:30.8345,(Bacth01Bb:26.6925,Flajo01Bb:26.5281):9.1256):21.4746):14.4231,((Gemob01Bo:40.5998,(Pl
ama01Bo:29.4117,(Blama01Bo:21.2192,Rhoba01Bo:28.5604):10.8598):8.7642):17.4646,(((Vicva01Bv:4
0.7170,Lenar01Bv:42.1258):13.5515,(Versp01Bv:46.7282,Opiba01Bv:45.0404):9.7755):8.3447,(CanPr01
Bv:31.3328,(Chlpn01Bv:13.0385,Chltr01Bv:13.8949):31.8389):27.0921):7.5969):8.0290):5.0101):6.7397,(
(((Theth01Bd:30.7799,Deira01Bd:38.2666):18.9921,(Rubxy01Ba:42.7450,(Biflo01Ba:42.3532,Myctu01Ba
:35.5812):20.5856):7.1378):8.3914,((Dehsp01Bh:45.1610,Chlau01Bh:37.0665):11.3927,(Glovi01Bc:27.3
297,(Proma01Bc:40.5533,((Acama01Bc:14.7899,Theel01Bc:16.7404):5.5897,(Synsp01Bc:20.6511,(Trier
01Bc:16.3717,(Nossp01Bc:0.2758,Anava01Bc:0.2406):10.3218):3.8772):4.8466):6.8180):6.4049):18.113
9):6.5438):6.2510,((Aquae01Bq:49.7884,(Ferno01Bt:30.2272,Thema01Bt:22.8174):24.8129):9.8613,((Fu
snu01Bu:43.8391,Mesfl01Bf:64.8579):9.1732,((Mooth01Bf:27.9007,Cloac01Bf:35.2075):6.7564,(Bacsu01
Bf:24.7567,Lacca01Bf:38.3900):13.8760):5.3383):7.1576):7.4726):6.3673):7.2299,((Solus01Bi:32.0055,A
ciba01Bi:30.4161):25.2051,((Sulsp02Bp:29.7525,Helpy01Bp:40.3668):39.5585,((Myxxa01Bp:39.8124,De
svu01Bp:46.1556):10.2976,((Ricpr01Bp:50.8076,(Metex01Bp:23.0949,Agrtu01Bp:22.8619):16.3410):15.6
247,((Neime01Bp:29.2603,(Metfl01Bp:20.3604,(Metpe01Bp:22.6169,Burma01Bp:16.1408):9.8286):5.144
9):7.5865,(Metca01Bp:26.5222,(Pseae01Bp:19.7872,Escco01Bp:23.2288):6.6808):7.9313):15.0439):8.3
809):4.2225):4.7923):7.6533):12.6121);

## Bootstrap analysis

The sets of 2294 pairs of $E^2_{MC,k}$ and $E^2_{UPM,k}$ values for each GT were sampled with replacement 2294 times. The $\Delta AIC$ values were computed using the sums of the sampled values. This procedure was repeated 1000 times.

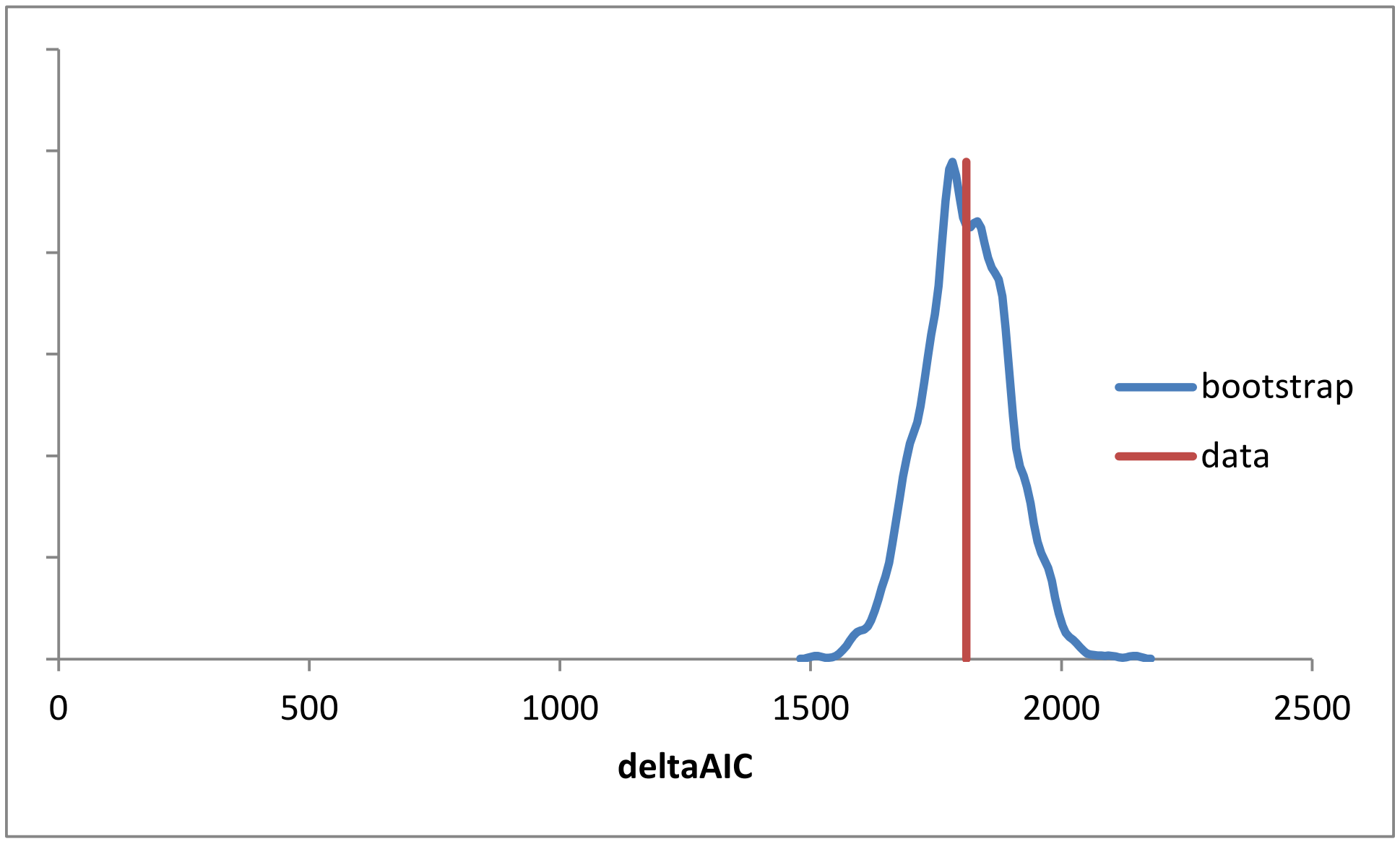

Distribution of the $\Delta AIC$ values for 1000 bootstrap samples (the curve was obtained by Gaussian-kernel smoothing of the individual data points). The red line indicates the $\Delta AIC$ value for the original set of GTs (1810.8).

The range of the $\Delta AIC$ values in the bootstrap sample is [1511.5, 2146.4]; 95% of the samples fell within the [1641.4, 1975.1] interval.

**Goodness of fit**

Analysis of the dependence of goodness of fit on other variables was performed using R as follows:

```
dat <- read.table("tmp.dat",header=TRUE,sep="\t")
```

Original data are transformed to a log scale:

```
dat$m <- log(dat$NM)                              # MAST GT size
dat$g <- log(dat$NG)                              # original GT size
dat$r <- log(dat$rUPM)
dat$cn <- log(dat$e2MC) - log(2*dat$NM-3)
dat$pn <- log(dat$e2UPM) - log(2*dat$NM-3)
dat$afit <- -(dat$cn+dat$pn)/2                    # average MC,UPM fit
                                                  # per MAST edge
dat$dfit <- log(dat$e2MC) - log(dat$e2UPM)        # difference between
                                                  # UPM and MC fit
dat$mg <- dat$m - dat$g                           # fraction of GT leaves
                                                  # retained in MAST

attach(dat)
```

We concentrate on what makes the individual gene trees to better fit the UPM model or the MC model. Thus, we apply linear analysis to explain the *dfit* value:

```
m00 <- lm(dfit ~ m+g+r+afit)
m01 <- step(m00)
summary(m01)

Call:
lm(formula = dfit ~ m + g + afit)

Residuals:
       Min         1Q     Median         3Q        Max
-0.3816085 -0.0323147 -0.0009259  0.0317164  0.3617230

Coefficients:
             Estimate Std. Error t value Pr(>|t|)
(Intercept)  0.010210   0.009105   1.121    0.262
m            0.047301   0.006432   7.354 2.67e-13 ***
g           -0.033069   0.004707  -7.026 2.80e-12 ***
afit         0.018875   0.001791  10.538  < 2e-16 ***
---
Signif. codes:  0 '***' 0.001 '**' 0.01 '*' 0.05 '.' 0.1 ' ' 1

Residual standard error: 0.05841 on 2290 degrees of freedom
Multiple R-squared: 0.09139,    Adjusted R-squared: 0.0902
F-statistic: 76.78 on 3 and 2290 DF,  p-value: < 2.2e-16
```

We find that *afit*, *m* and *g*  variables significantly affect *dfit*, whereas *r* doesn't. Since *m* (MAST GT size) and *g* (original GT size) have coefficients with opposite signs and similar absolute values, we further tested the hypothesis that it is the fraction of original GT leaves retained in MAST (*mg*) that is important:

```
m10 <- lm(dfit ~ mg+g+r+afit)
m11 <- step(m10)
summary(m11)

Call:
lm(formula = dfit ~ mg + g + afit)

Residuals:
       Min         1Q     Median         3Q        Max 
-0.3816085 -0.0323147 -0.0009259  0.0317164  0.3617230 

Coefficients:
            Estimate Std. Error t value Pr(>|t|)    
(Intercept) 0.010210   0.009105   1.121    0.262    
mg          0.047301   0.006432   7.354 2.67e-13 ***
g           0.014232   0.003244   4.387 1.20e-05 ***
afit        0.018875   0.001791  10.538  < 2e-16 ***
---
Signif. codes:  0 '***' 0.001 '**' 0.01 '*' 0.05 '.' 0.1 ' ' 1 

Residual standard error: 0.05841 on 2290 degrees of freedom
Multiple R-squared: 0.09139,    Adjusted R-squared: 0.0902 
F-statistic: 76.78 on 3 and 2290 DF,  p-value: < 2.2e-16
```

In the presence of *mg*, *g* remains significant and positively correlated with *dfit*. Thus, we conclude that the trees that fit UMP better than MC tend to:

- retain higher fraction of the original GT leaves in MAST (these are the trees that are least affected by HGT and tree reconstruction artifacts)
- have wider distribution in the prokaryotic world
- show a better fit to ST on average (these are the trees with the least deviations from any supertree)

Finally, to determine whether functional characteristics of the gene family play a role (*Class* variable), we used the following procedure:

```
r20 <- lm(dfit ~ mg+g+afit+Class)
r21 <- step(r20)
summary(r21)

Call:
lm(formula = dfit ~ mg + g + afit)

Residuals:
       Min         1Q     Median         3Q        Max 
-0.3816085 -0.0323147 -0.0009259  0.0317164  0.3617230 

Coefficients:
            Estimate Std. Error t value Pr(>|t|)    
(Intercept) 0.010210   0.009105   1.121    0.262    
```

```
mg           0.047301   0.006432   7.354 2.67e-13 ***
g            0.014232   0.003244   4.387 1.20e-05 ***
afit         0.018875   0.001791  10.538  < 2e-16 ***
---
Signif. codes:  0 '***' 0.001 '**' 0.01 '*' 0.05 '.' 0.1 ' ' 1

Residual standard error: 0.05841 on 2290 degrees of freedom
Multiple R-squared: 0.09139,    Adjusted R-squared: 0.0902
F-statistic: 76.78 on 3 and 2290 DF,  p-value: < 2.2e-16
```

The *Class* variable was excluded from the model in the course of the stepwise reduction. Consequently, we find that the functional assignment of the gene family is unimportant.

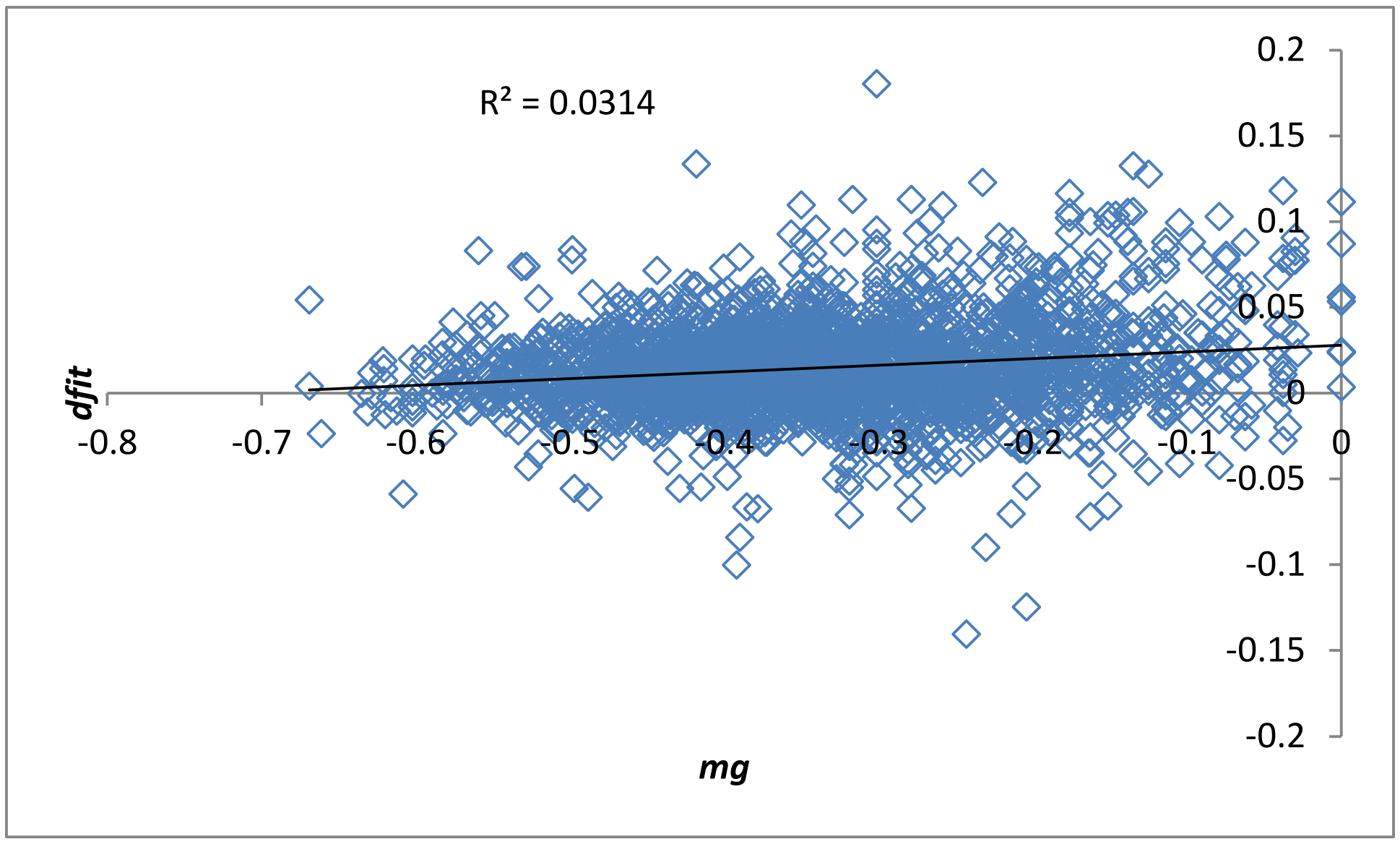

Relative goodness of fit for the UPM vs the MC model (*dfit*) plotted against the fraction of original GT leaves retained in MAST (*mg*).

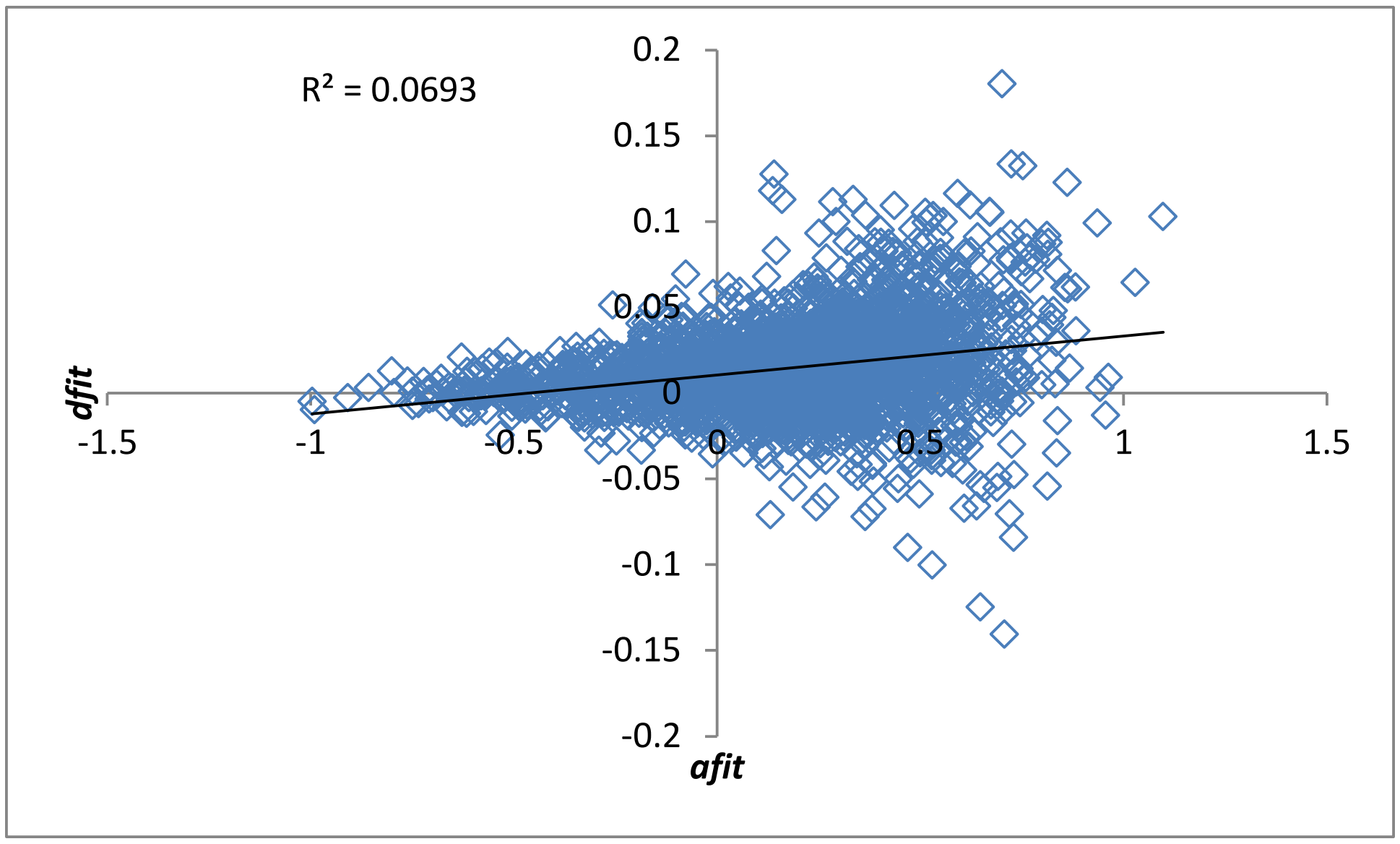

Relative goodness of fit for the UPM vs the MC model (*dfit*) plotted against the average goodness of fit (*afit*).

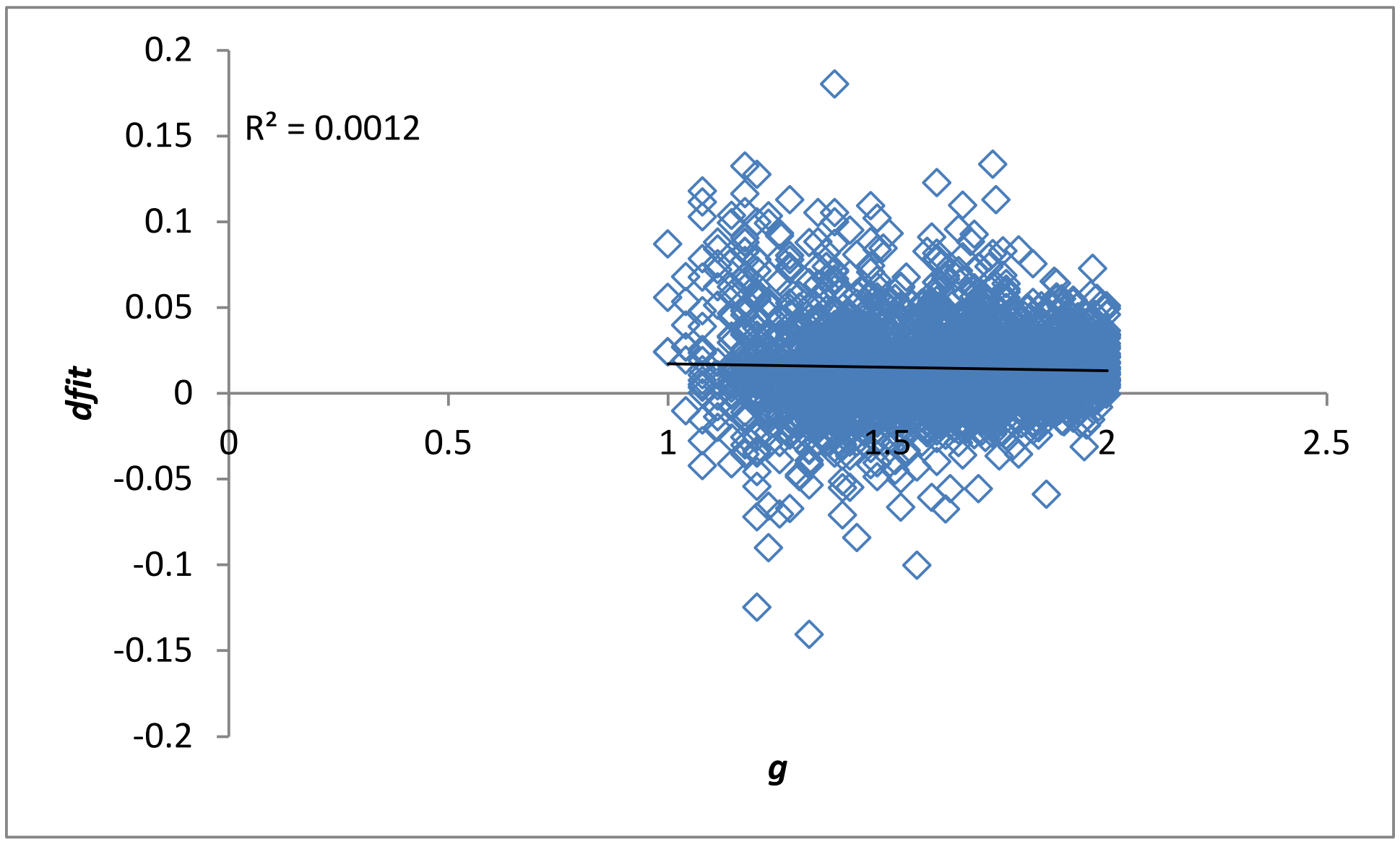

Relative goodness of fit for the UPM vs the MC model (*dfit*) plotted against the original GT size (*g*).